# Modeling of hydrogen and hydroxyl group migration on graphene

D.W. Boukhvalov

*Computational Materials Science Center, National Institute for Materials Science,*

*1-2-1 Sengen, Tsukuba, Ibaraki 305-0047, Japan*

*Density functional calculations of optimized geometries for the migration of single hydrogen and hydroxyl groups on graphene are performed. It is shown that the migration energy barrier for the hydroxyl group is three times larger than for hydrogen. Crucial role of supercell size for the values of the migration barriers are discussed. The paired migration of hydrogen and hydroxyl groups has also been examined. It could be concluded that hydroxyl groups based magnetism is rather stable in contrast with unstable hydrogen based magnetism of functionalized graphene. The role of water in the migration of hydroxyl groups is also discussed, with the results of the calculations predicting that the presence of water weakens the covalent bonds and makes these groups more fluid. Increasing of number of water molecules associated with hydroxyl group provides grown of the migration energy.*

E-mail: D.Bukhvalov@science.ru.nl

# 1. Introduction

Graphene functionalization is one of the routes for manipulation of its chemical and physical properties [1]. Transformation of graphene to graphane [2] has shown new frontiers in carbon chemistry. Modeling realistic mechanisms of graphene functionalization is an important issue in modern computational physics. Previous research on graphene hydrogenation has considered the stability of different hydrogen configurations (see Ref. [1] and references therein) or the pathways for hydrogen adsorption on to graphene [3, 4]. Potential barriers for the migration of hydrogen on graphene have only been discussed for a few possible pathways [5-9]. This requires more detailed and systematic studies of the adatoms and functional groups on graphene.

Reduced graphene oxide (also known as functionalized graphene [10-12]) is a graphene sheet functionalized by hydroxyl groups with coverage of about 10% [13]. Knowledge about the migration of these groups is important for possible further functionalization or reduction of this compound. Hydrophilicity is the key property of graphene oxide as it opens a way for the application of graphene oxide as a scaffold for biomolecules deposition [14]. Recent intriguing experimental reports about enormous graphene oxide behavior under water [15, 16] and ethanol [17] pressure require a study of the hydroxyl group migration in the presence of water molecules. Several interesting models developed for hydroxyl groups [18] and hydrogen [19, 20] randomly distributed on graphene surface also make this modeling prompt.

In the present work I have been examine the potential barriers for the migration of single hydrogen and hydroxyl groups on a graphene surface (see inset in Fig. 1a) and the migration of one functional group in pair from a less to a more energetically favorable configuration (see Fig. 1b,c). For hydroxyl groups the role of water molecules in migration is examined.

**2 Computational method and used models**

The modeling was carried out by density functional theory (DFT) realized in the pseudopotential code SIESTA [21], as was done in our previous works [1, 13, 22]. All calculations are done using the generalized gradient approximation (GGA-PBE) [23] which is more suitable for the description of graphene-adatom chemical bonds [22] and systems with few hydrogen bonds [24]. Full optimization of all atomic positions was performed. For careful modeling of the migration process I chose $8\times8\times1$ supercell containing 128 carbon atoms in the graphene sheet. Thus the large size supercell guarantees the absence of any overlap between the distorted areas around chemisorbed groups [1]. The lattice distortions after chemisorption of hydrogen and hydroxyl groups are the same as in our previous works [1, 13]. For the carbon atoms remote from the chemisorbed group only neglected deviations from ideal carbon-carbon distance in graphene has been found. All calculations were carried out for an energy mesh cut off of 360 Ry and a k-point mesh $6\times6\times1$ in the Mokhorst-Park scheme [25].

Standard method that has using now for energy barriers computation is nudged elastic bands (NEB) method [26, 27]. This method is rather general and valid for migration processes, dissociation and chemical reactions. NEB method requires large computational costs and not implemented to the standard DFT codes. For migration of adatoms over graphene we exactly know atomic structure and energetics for initial and final point of the process. Thus the simplified algorithm within NEB method could be provided for studied systems.

First step of calculation energy barriers of the migration is calculation of atomic structure and total energy for initial point of the migration. This step are identical the calculation of single adatom or specie chemisorbed on carbon atom of graphene lattice. For the study of the migration in pair the calculation for initial and final combinations of atoms is required. Second step of calculation is the check of possible routes of migration. For exploration of the possible pathways require to put the chemisorbed specie over the center of C-C bond at the height obtained on the first step of calculation and perform the full optimization of atomic structure of whole system with fixed position only the nearest to carbon substrate atom of chemical specie (hydrogen for the case of hydrogen migration and oxygen for the case of hydroxyl group migration). Furthermore for examination of the possible alternative pathways need to shift the specie up and down from the point discussed above about 0.1 Å and perform the described above calculations. If the first point of this step of calculations does not correspond with minimal total energy further varying of height of the chemical specie are required until the energy minimum for the intermediate position will be obtained. Last step of the calculation correspond with division of the pathway of migration to ten equal segments and calculation of the total energy of system for each step with taking into account full optimization of atomic structure except nearest to graphene atom of the studied chemical specie. If the distance between graphene and chemical specie is not the same for initial and final or intermediate steps of migrations smooth changes of this parameters for each step of migration are required. The difference between the total energy of the system at initial step of migration and the highest energy for the intermediate steps of migration defined as the value of energy barrier.

**3 Migration of single hydrogen and hydroxyl groups without water**

First I had studying the migration of a single hydrogen adatom on the graphene surface using the above proposed scheme. Additionally, to check possible changes in the distance between the hydrogen and graphene, I calculated the total energies for different distances between the graphene and the hydrogen atom. It was found that the lowest energy corresponds to the same distance between hydrogen and graphene sheet as in the stable final and start positions (1.57 Å). It needs to note that large size of used supercell provides insignificant changes of atomic structure of flat graphene around distortions near migration ways. As we can see in Fig. 1a the total energy of the system increases smoothly during migration from the starting to the middle point and then smoothly decreases for the second part of the migration. The energy barrier for single hydrogen migration is 0.29 eV which corresponds to the experimentally observed relative stability of hydrogen adatoms on graphene [29].

Calculated value of hydrogen migration is smaller that achieved in previous calculations [5-9]. For verify used method and choice of technical parameters I have been performed calculations for different sizes of supercell. For the supercells where single hydrogen adatoms separated less than 6 lattice parameters (about 1.5 nm) migration barrier is about 1 eV (see Fig. 2) that is in perfect agreement with previous theoretical studies [5, 7]. Further separation of adatoms provides decay of migration barrier. The cause of founded phenomenon is very wide (about 1 nm radius) propagation of unpaired electron and lattice distortion near chemisorbed hydrogen [1]. Two hydrogen adatoms disposed closed than 2 nm is not independent enough. For non-interactive adatoms all directions of migrations is equivalent but for the case of weakly interacting adatoms appear special and very energetically favorable (see below) direction of migration (see inset of Fig. 2). Instead of migration of one of hydrogens to another sublattice near to hydrogen we move both adatoms in much less favorable direction. It provide large grow of migration barrier for the case of small sizes of supercell.

For the migration of a single hydroxyl group on graphene the situation is quite different. In contrast with hydrogen this group slightly stretches away from graphene flat area from 1.974 to 2.181 Å. A step by step increase of the distance from this group to the graphene sheet is required. However, the main difference from the hydrogen case is in the energetics of migration. The first small step from the equilibrium position corresponds to an energy barrier of 0.94 eV. This value is three times larger than that for hydrogen because the binding energy between the hydroxyl group and graphene is also about three times larger [11, 18]. Further migration of the hydroxyl group overcomes the energy barrier of less than 0.1 eV. This small value causes a large distance between the moved hydroxyl groups and the graphene sheet. Calculated values are much higher than typical values of weak bonds (below 50 meV) and special corrections for taking into account van der Walls forces is not required. The reciprocal orientation of distant hydroxyl groups noted in our previous works [13, 30] provides for the high concentration of this impurities formation of hydroxyl-hydroxyl web and diminishment of discussed for the hydrogen supercell size effects. Further grow the size of supercell results the increase of the migration barriers values (see Fig. 2).

**4 Migration barriers in the pairs and stability of magnetic configuration**

For modeling of the migration of the hydrogen and hydroxyl groups inside the pair two final configurations were chosen. The first configuration (Fig. 1b) corresponds to two groups from different sides of the graphene sheet on ortho positions on the hexagon, and the second configuration is more stable where the groups are in the para positions of the hexagon on the same side of the graphene (Fig. 1c). The starting point is chosen up to the next carbon atom. In contrast with a single adatom or group migration where the starting and final configuration are equivalent, now for both (hydrogen and hydroxyl groups) the final configuration is energetically more favorable. For detailed discussions about the nature of this phenomenon see Ref. [1, 23]. It is necessary to note that in both cases the starting configuration is ferromagnetic [28] and the final nonmagnetic [23].

The above described energy difference between the starting and final points of migration completely change the energetics of the process (see Fig. 1b, c). When hydrogen or hydroxyl groups overcome the energy barriers corresponding to the initial stages of the breaking of covalent bonds they move to the final configuration without any further barriers. It should be noted that for hydrogen the described barrier is 0.07 eV for two side hydrogenation (Fig. 1b) and 0.26 eV for one side hydrogenation (Fig. 1c), corresponding with the experimentally observed presence of hydrogen on graphite only in the most stable configurations [6] and the pairing of hydrogens adatoms on graphene [29]. For hydroxyl groups these barriers are about 0.8 eV for both cases. It means that in contrast with unstable hydrogen based magnetism the magnetic configurations built from hydroxyl groups will be rather stable. Discussed stability of magnetism is corresponding to the survival of unsaturated dangling bonds. Exact description of single adatom magnetism requires taking into account self-interaction correction scheme [31]. The experimental observation of ferromagnetism in annealed graphene oxide [32] also supports this assertion.

**5 Migration of hydroxyl groups in the presence of water**

Previous theoretical studies adsorption of water at the large polyaromatic hydrocarbons [33-35], on the edges of mentioned molecules [36] or at the pure infinite graphene surface [37] has been considered. In all described cases water molecules are bind with graphene by the weak van der Walls bonds, which is correspond to the hydrophobic properties of non-functionalized carbon nanosystems. Hydrophilic properties of graphene oxide caused the formation of hydrogen bonds between water and hydroxyl groups were deeply studied mainly experimentally [11, 38, 39].

For the modeling I have used as initial model single water molecule connected by hydrogen bond with hydroxyl group chemisorbed on graphene (Fig. 3a). The binding energy of this bond varies from 0.42 eV for hydroxyl groups in the para position (Fig. 1c and Fig. 2) to 0.53 eV for a single hydroxyl group (Fig. 1a) and a pair in the ortho position (Fig. 1b). Calculated values evidence about the hydrophilisity of studied compound in agreement with experimental results [11, 38, 39]. The distance between hydrogen atom of hydroxyl group and oxygen atom of the single water molecule vary very insignificant from 1.88 to 1.91 Å. Calculated values of distances and binding energies are reasonable for the hydrogen bonds [40]. Decay of number of electrons in single water molecule on pure graphene is 0.004. In the presence of the hydroxyl group the electron transfer from water to graphene is increased to 0.31 electrons per water molecule. Thus injections of electrons provide a partial compensation of unpaired electrons that are formed in the case of the covalent bond breaking and correspond to the weakening of the covalent bond without significant changes in the electronic structure of functionalized graphene (Fig. 4). The presence of water molecules also enhances the distortion of the graphene sheet (see as example Fig. 3 and 5) and the initial deflection of hydroxyl groups that provide shortening of the migration pathway (Fig. 1). The height of the carbon atom bonded with hydroxyl group over graphene flat increase from 0.49 Å for single hydroxyl group to 0.66 Å for the same group associated single water molecule

These rich doping and closeness of hydroxyl water binding energies to the energies required for first steps from the stable configuration provide dramatic changes in the picture of migration. In the presence of water the energetics of the migration of single hydroxyl groups are close to those of single hydrogen. In the case of pairs the energy barriers disappear and hydroxyl groups freely migrate to more energetically favorable final configurations resulting in the destabilization of the hydroxyl group based magnetism. Other effect of the hydrogen bonds formation is reduction of the role of distance between single hydroxyl groups to migration barrier (Fig. 2) due to destroy the reciprocal orientation of hydroxyl groups described above.

Previous experimental studies [13] suggest for the dependence of graphene oxide structural properties from the water concentration. For the study of this phenomenon the calculations of interactions of small water clusters with single hydroxyl group have been performed. The numbers of the water molecules in the considered clusters have been chosen three (one water connected by hydrogen bond with hydroxyl groups and two other molecules connected with the hydrogen atoms of this molecule) and five. Results of calculations suggest for the significant increase of water – hydroxyl group bond lengths (see Fig. 3). The shapes of the small water cluster and distances between water molecules are different from the similar water clusters on non-functionalized graphene [33, 34]. The cause of diversity between pure and functionalized graphene case are presence of additional hydroxyl group connected with water molecule from the clusters and drain the charge from water molecules to carbon substrate. Increasing of the number of the water molecules significantly reduce of the charge transfer to graphene from 0.31 electrons for single water molecule to 0.162 and 0.112 electrons for three and five molecules in water clusters respectively. The shift up of the carbon atom connected with hydrogen is not significantly depending from the number of water molecules (0.75 and 0.73 Å for three and five molecules in cluster respectively). But the radius of the graphene flat corrugations is not depend form the number of water molecules and keep the same value (10 Å) as for the case of single hydrogen atom or hydroxyl group [1]. Diminishment of the electron doping of graphene, enormous local distortion of graphene sheet and growing up the number of water molecules connected with single hydroxyl group provides significant increase of the migration barrier from 0.58 eV for single water molecule to 1.07 and 3.32 eV for the larger studied water clusters. The raise of the concentration of water leads decay of hydroxyl groups on graphene fluidity.

## 6 Conclusions

Using DFT modeling and the proposed new algorithm for calculation of the energetics of adatom and functional groups on graphene I have shown that (i) single hydrogen atoms on graphene have an approximately three times smaller energy barrier for migration than hydroxyl groups; (ii) in the case of pairs the migration of hydrogen also requires much smaller energy than for hydroxyl groups; (iii) in contrast with the unstable hydrogen based magnetism hydroxyl groups could be a source for magnetic graphene; (iv) the presence of water strongly weaken hydroxyl group graphene covalent bonds making these groups very fluid on the graphene surface. The last result could be the key for understanding the anomalous graphene oxide expansion under water pressure [15] and may form a basis for new methods of graphene oxide total reduction [8].

**Acknowledgements** I gratefully acknowledge G. N. Newton for careful reading the manuscript.

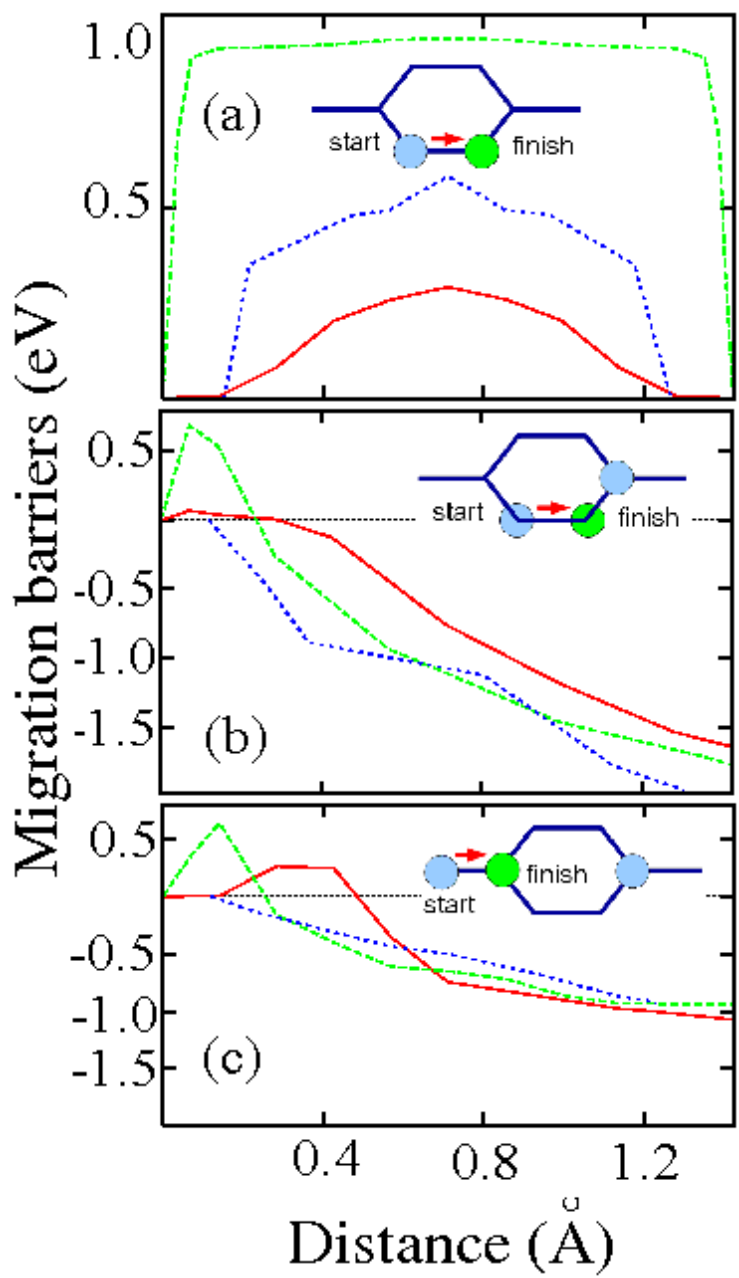

**Figure 1** Energy barriers for the migration of hydrogens (solid lines), hydroxyl groups (dashed light lines) and hydroxyl groups bonded with water (dotted dark line) for different types of migration; a) Migration of a single adatom or group, b) migration to the pair in the ortho position from both sides of the graphene sheet, c) migration to the para position from one side of the graphene sheet. On the insets of the panels reciprocal sketches for the migration pathways are presented.

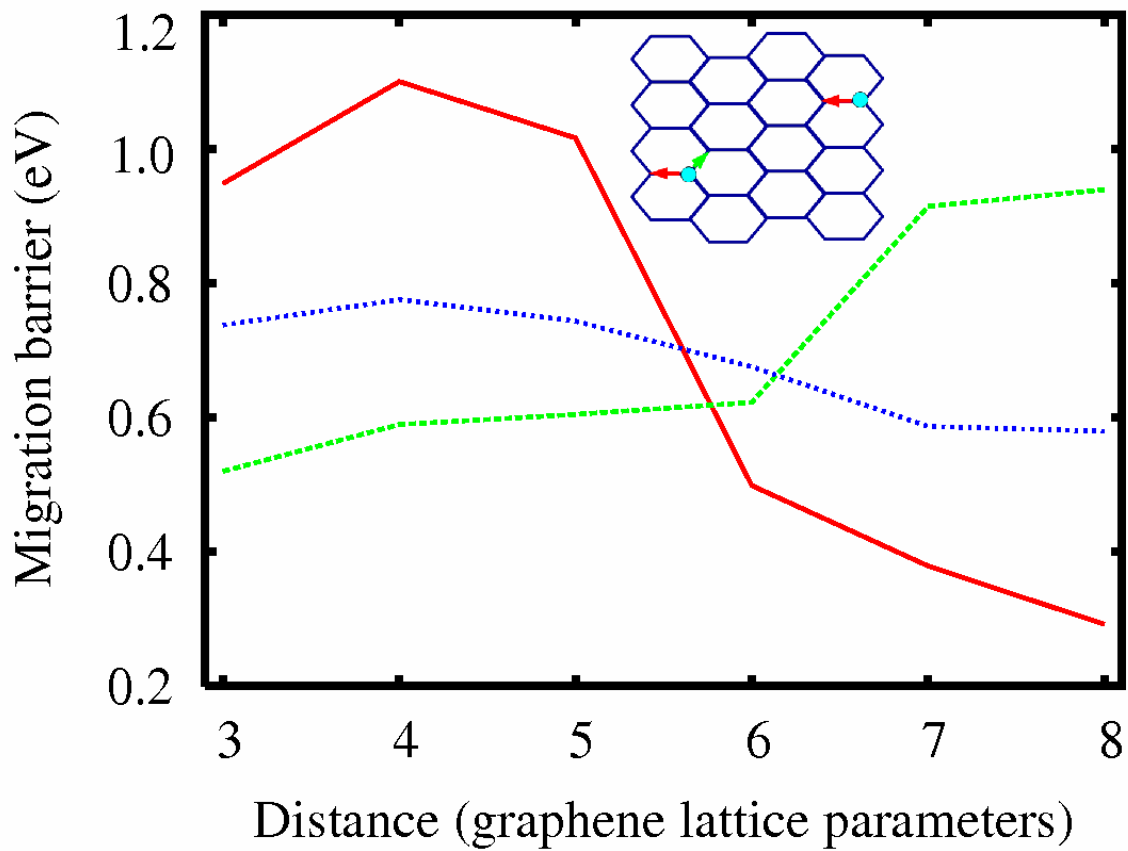

**Figure 2** Migration barriers of single hydrogen migration as function of distance between equivalent adatoms for the single hydrogen adatom (solid red line), hydroxyl group (dashed light line) and hydroxyl group with attached water molecule (dark dotted line). On inset a sketch two far placed hydrogen adatoms (light circles) on graphene lattice. By darker arrows is shown modeled migration, by lighter arrow – most energetically favorable.

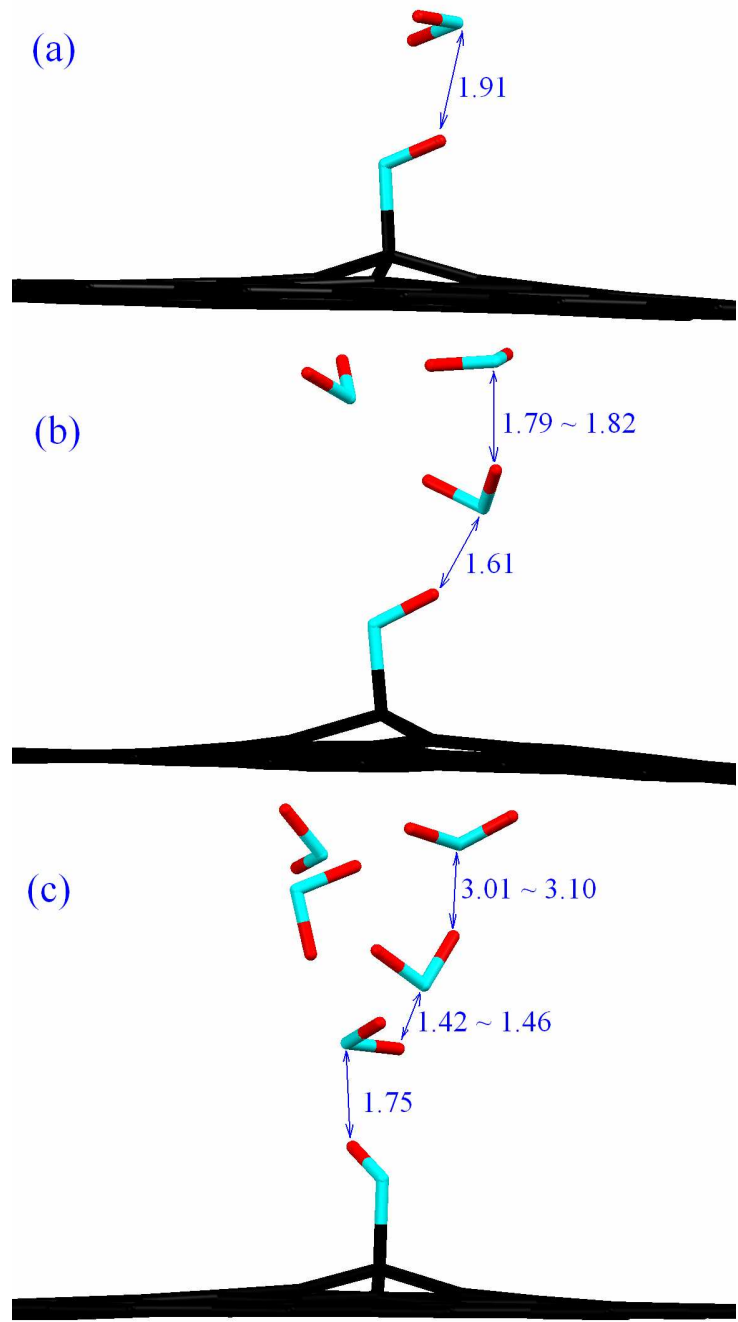

**Figure 3** Optimized atomic structure of single hydroxyl group (oxygen atom are shown by light gray, hydrogen by dark gray) chemisorbed on graphene scaffold (shown by black) in the presence of one (a), three (b) and five (c) water molecules. All values of the hydrogen bonds are measured in Angstroms.

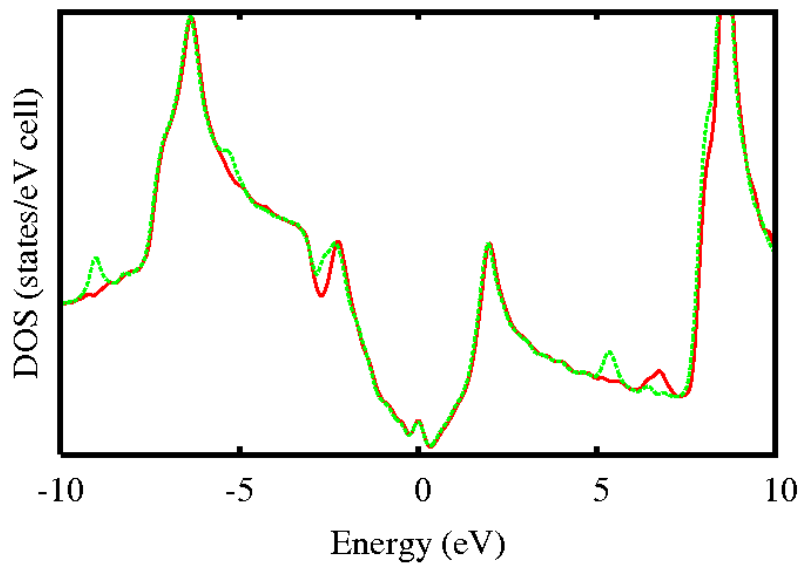

**Figure 4** Total densities of states for single hydroxyl group on graphene in absence (solid dark line) and presence (dashed light lie) of water molecule.

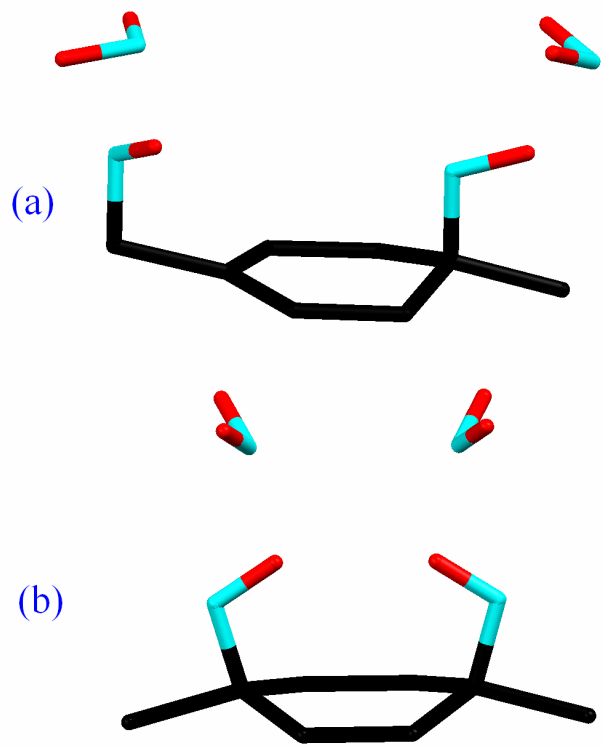

**Figure 5** Optimized atomic structures for initial (a) and final (b) steps of migration of hydroxyl group in para position (see Fig. 1c) in the presence of water molecules.